\documentclass[aps,twocolumn,superscriptaddress,nofootinbib]{revtex4}
\usepackage{graphicx}
\usepackage{dcolumn}
\usepackage{epsfig}
\usepackage{amsmath}
\usepackage{amssymb}

\begin{document}

\newcommand{\alumina}[0]{Al$_2$O$_3$}
\newcommand{\Kalumina}[0]{$\kappa$-Al$_2$O$_3$}
\newcommand{\Aalumina}[0]{$\alpha$-Al$_2$O$_3$}


\title{Computational  scheme for \textit{ab-initio} predictions of chemical compositions
interfaces realized by deposition growth}

\author{Jochen Rohrer}
\email{rohrer@chalmers.se}
\affiliation{%
BioNano Systems Laboratory,
Department of Microtechnology,
MC2,
Chalmers University of Technology,
SE-412 96 Gothenburg
}%
\author{Per Hyldgaard}%
\affiliation{%
BioNano Systems Laboratory,
Department of Microtechnology,
MC2,
Chalmers University of Technology,
SE-412 96 Gothenburg
}%

\begin{abstract}
We present a novel computational scheme
to predict chemical compositions at interfaces
as they emerge in a growth process.
The scheme uses the Gibbs free energy of reaction associated
with the formation of interfaces with a specific composition
as predictor for their prevalence.
It explicitly accounts  for the growth conditions 
by rate-equation modeling of the deposition environment.
The Bell-Evans-Polanyi principle motivates our emphasis on an
effective nonequilibrium thermodynamic description
inspired by chemical reaction theory.
We illustrate the scheme by characterizing the interface
between TiC and alumina.
Equilibrium thermodynamics favors a nonbinding
interface, being in conflict with the wear-resistant nature
of TiC/alumina multilayer coatings.
Our novel scheme predicts that deposition of a strongly adhering
interface is favored under realistic conditions.
\end{abstract}


\maketitle

\section{Introduction}
\label{Intro}
Density functional theory calculations are today routinely applied
to characterize structural and electronic properties of condensed matter systems.
They serve as an important complement to and extension of experimental methods
\cite{complement}.
Atomic (and electronic) structure and chemical composition are inseparably interwoven.
Information about chemical composition at, for example, interfaces (including surfaces)
is therefore of great importance for reliability of such calculations.

Traditional \textit{ab initio} atomistic thermodynamics methods 
aim at describing and predicting compositions at oxide surfaces
\cite{PhysRevB.62.4698,AIT_Scheffler}.
These schemes assume that equilibrium is established between the oxide and surrounding O$_2$.
This criterion is often justified for oxide surfaces
that are in direct contact with an O-rich environment.
For (solid-solid) interfaces, the situation is more complicated.
Interfaces are typically exposed to the surrounding environment
only at the moment of creation.\footnote{
We assume that the interface is created by a deposition
process from a gas phase.}
Furthermore, oxides are seldomly grown directly from O$_2$,
and it is not clear what gas(es) they are in 
(dynamic) equilibrium with during deposition, if at all.
A generalization of the \textit{ab initio} atomistic thermodynamics
scheme from surfaces to interfaces \cite{PhysRevB.70.024103} 
can therefore be problematic.
In fact, we have shown \cite{AluminaStructureSearch}
that the equilibrium configuration 
at the interface between TiC and alumina
predicted by such a generalized scheme
does not describe the wear-resistance
of TiC/alumina multilayer coatings \cite{Halvarsson1993177}.
We have attributed this inconsistency to the fact that the scheme 
does not account for the actual growth conditions. 
This  may  be a serious shortcoming also for other interfaces realized by deposition growth.

This paper presents a computational scheme that 
explicitly accounts for deposition conditions.
At the same time no \textit{a priori} equilibrium assumptions
are introduced.
The scheme is therefore capable to predict chemical compositions 
at interfaces (including surfaces) as they arise in a deposition process.
While kinetic accounts are generally required to describe the full 
nonequilibrium nature of growth,
the Bell-Evans-Polanyi (BEP) principle \cite{BEP}
motivates an effective thermodynamic description
inspired by chemical reaction theory.

The key elements of our scheme are the use of Gibbs free energies of reaction
as a predictor for the chemical composition
and a modeling of the deposition conditions
in terms of rate equations describing the deposition environment.
Our approach extends the method of \textit{ab initio} thermodynamics
of deposition growth for surfaces \cite{AIT-DG_TiX}
to interface modeling.
Applying the scheme to the TiC/alumina interface
in a chemical vapor deposition (CVD) environment
we predict the formation of strongly adhering structures
in agreement with the wear-resistance  \cite{Halvarsson1993177}
of CVD TiC/alumina multilayers.

The paper is organized as follows.
Section~\ref{strategy} motivates the use of the Gibbs free energy of reaction 
as a predictor of chemical composition.
Section~\ref{modeling} contains the details of our modeling.
Results are presented in Sec.~\ref{results} 
and Sec.~\ref{conclusions} contains our conclusions.

\section{Predictor for as-grown chemical composition}
We use the Gibbs free energies of reaction $G_r$
as a predictor for the prevalent chemical composition
at surfaces and interfaces realized by deposition growth.
We justify the use of $G_r$ as predictor using 
chemical reaction theory \cite{Demirel,Tschoegl}
and the BEP principle \cite{BEP}.

A description of nonequilibrium growth generally requires a
kinetic description.
The prevalent surface or interface composition 
will in principle depend on the reaction barriers for 
all relevant processes.
However, the BEP principle \cite{BEP}
motivates our emphasis on an effective nonequilibrium thermodynamic approach.
The BEP principle states that the more exothermic
a reaction is,
the smaller is the  associated reaction barrier.
The nonequilibrium system will therefore continuously evolve
along paths that release the largest Gibbs free energy of reaction. 
In some cases we can even obtain quantitative predictions
for the prevalence of competing chemical compositions
without having to calculate specific rates.

The Gibbs free energy of reaction is defined as the gain in Gibbs free energy
in a chemical reaction.
For a  general reaction, according to chemical reaction theory, 
$G_r$ is related to the forward ($f$) and backward ($b$) reaction rates
$\Gamma_f = k_f \Pi_r[\text{R}_r]^{\nu_r}$ and $\Gamma_b = k_b \Pi_p[\text{P}_p]^{\nu_p}$
via 
\begin{align}
-\beta G_r = \ln \Gamma_f/\Gamma_b
\label{eq:GrK}
\end{align}
Here $\beta$ is the inverse temperature in units of energy,
$k_{\{f,b\}}$ is forward and backward rate constant,
$[\text{R}_r]$ ($[\text{P}_p]$) is the concentration
of the $r$-th ($p$-th) reactant R (product P),
and $\nu_ {\{p,r\}}$ is the corresponding 
stoichiometric coefficient.\footnote{
In this section we count stoichiometric coefficients positively,
opposed to standard chemical nomenclature,
where stoichiometric coefficients of reactants 
are counted negatively.}
The key observation is that the ratio of forward and backward rates 
in a given chemical process is entirely given by the nonequilibrium 
Gibbs free energy of reaction.
This quantity can be evaluated without knowing the kinetic details
of the process.

\begin{figure}
\epsfig{file=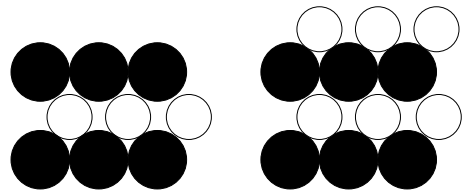, width=6cm}
\caption{
\label{fig:AB}
Illustration of two different terminations
at the surface of a binary material AB.}
\end{figure}

To illustrate the approach,
we first discuss the problem of surface termination in deposition growth \cite{AIT-DG_TiX},
seeking to predict which surface (A terminated or B terminated)
will emerge when growing a binary material AB,
see Fig.~\ref{fig:AB}.
We consider two coupled reactions $I$ and $II$ that change the surface termination
from A to B and vice versa,
\begin{subequations}
\begin{align}
I:~\text{B}+\text{React}_I
\mathop{\rightleftharpoons}^{\Gamma_f^I}_{\Gamma_b^I} 
\text{A}+\text{Prod}_I \\
II:~\text{A}+\text{React}_{II}
\mathop{\rightleftharpoons}^{\Gamma_f^{II}}_{\Gamma_b^{II}} 
\text{B}+\text{Prod}_{II}.
\end{align}
\label{eq:Reactions}
\end{subequations}
Here, $\text{React}_{\{I,II\}}$ and $\text{Prod}_{\{I,II\}}$ collectively denote the
reactants and products in the two reactions.
The forward and backward rates are the same as in (\ref{eq:GrK}) but have an additional index
to differentiate between the two reactions.
Taking the difference between the free energies of reaction in $I$ and $II$,
we find after exponentiation and use of the geometric mean $\langle x,y\rangle_{\text{gm}}=(xy)^{-1/2}$
\begin{align}
\exp\left(-\beta[G_r^I-G_r^{II}]\right)=\frac{\Gamma_f^I\Gamma_b^{II}}{\Gamma_b^I\Gamma_f^{II}}
=\left[\frac{\langle\Gamma_f^I,\Gamma_b^{II}\rangle_{\text{gm}}}{\langle\Gamma_b^I,\Gamma_f^{II}\rangle_{\text{gm}}}\right]^2.
\label{eq:ExpGr}
\end{align}

\label{strategy}

The reactions  in (\ref{eq:Reactions}) can also 
be  described by the rate equations,
\begin{subequations}
\begin{align}
\partial_t P_{\text{A}}=
-(\Gamma_b^{I}+\Gamma_f^{II}) P_{\text{A}} 
+(\Gamma_f^{I}+\Gamma_b^{II}) P_{\text{B}}\\
\partial_t P_{\text{B}}=
(\Gamma_b^{I}+\Gamma_f^{II}) P_{\text{A}} 
-(\Gamma_f^{I}+\Gamma_b^{II}) P_{\text{B}}.
\end{align}
\end{subequations}
Here we have introduced  the probability $P_{\text{\{A,B\}}}$ for 
observing either an A or B terminated surface.
Using the arithmetic mean $\langle x,y\rangle_{\text{am}}=(x+y)/2$,
the steady-state solution provides a ratio between these probabilities,
\begin{align}
\frac{P_{\text{A}}}{P_{\text{B}}}=
\frac{(\Gamma_f^{I}+\Gamma_b^{II})}{(\Gamma_b^{I}+\Gamma_f^{II})}=
\frac{\langle\Gamma_f^{I},\Gamma_b^{II}\rangle_{\text{am}}}{\langle\Gamma_b^{I},\Gamma_f^{II}\rangle_{\text{am}}}.
\label{eq:P}
\end{align}

In dynamic equilibrium \cite{Demirel,Tschoegl},
being characterized by $G_r^I+G_r^{II}=0$ or equivalently 
$\Gamma_f^{I}/\Gamma_b^{II}=\Gamma_b^{I}/\Gamma_f^{II}$,
we find by combining (\ref{eq:ExpGr}) and (\ref{eq:P}) that
\begin{align}
\frac{P_{\text{A}}}{P_{\text{B}}}\big|_{\text{dyn. eq.}}=\exp\left(-\beta[G_r^I-G_r^{II}]/2\right).
\label{eq:Pratio}
\end{align}
In particular, if the reaction that creates an A-terminated surface
has lower $G_r$ than the reaction that creates a B-terminated surface,
here $G_r^{I}<G_r^{II}$, an A-terminated surface is more likely.
Away from dynamic equilibrium, this relation is no longer exact.
However, comparison of $G_r$
for different terminations still indicates which 
termination is most prevalent,
since the geometric and arithmetic means 
[Eqs.   (\ref{eq:ExpGr}) and  (\ref{eq:P})]   
are approximately equal, Ref.~\cite{AIT-DG_TiX}.

By analogy to the case of surface terminations,
we  can also use differences in $G_r$, see (\ref{eq:Pratio}), 
as measure to predict chemical compositions at interfaces formed by deposition growth
(and assumed to have retained the structure specified by the deposition environment).
Considering related interface configurations of similar thickness but
different compositions,
the one with lowest (most negative) Gibbs free energy of reaction
is most likely to describe the nature of the interface as it is formed 
by deposition growth.
We note that $G_r$ will in general decrease (if the reaction is favorable)
as the number of constituents in the film describing the interface
increases by any integer multiple of the bulk stoichiometry.
The predictor is useful when comparing films of similar thickness.
Again it is possible to obtain quantitative predictions without having to 
consider kinetic details of the deposition, see also Ref.~\cite{AluminaAdhesion}.

\section{Modeling}
\label{modeling}

\begin{figure}
\epsfig{file=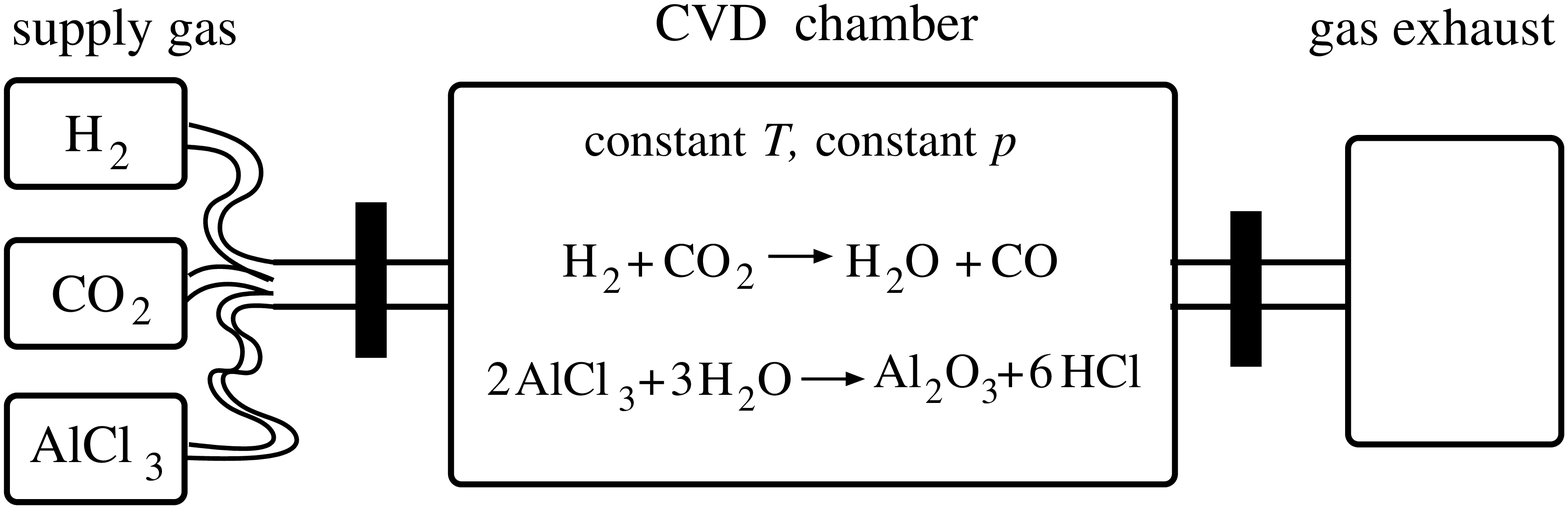, width=8.5cm}
\caption{
\label{fig:CVD}
Illustration of chemical vapor deposition of alumina.
A H$_2$-AlCl$_3$-CO$_2$ gas mixture is supplied
to a chamber at rate $R_S$.
The chamber is kept at fixed temperature $T$ and fixed
total pressure $p$.
The gases react to simultaneously form water and alumina.
Reaction products and unused reactants are exhausted
at rate $R_E$.
}
\end{figure}

\subsection{Materials Background}
We illustrate our computational scheme by studying
the interface composition between TiC and alumina.
TiC/alumina multilayers are commonly used as wear-resistant coating on 
cemented-carbide cutting tools \cite{Halvarsson1993177}.
They are routinely fabricated by chemical vapor deposition (CVD).

Figure \ref{fig:CVD} illustrates the experimental setup for CVD of alumina \cite{Ruppi200150}.
A H$_2$-AlCl$_3$-CO$_2$  supply gas mixture is injected into a hot chamber 
which is kept at a fixed temperature and a fixed total pressure.
The CVD process proceeds in two steps which, however, take place in parallel.
Water forms at an (unknown) rate $R_{\text{H$_2$O}}$  according to 
\begin{equation}
\text{H$_2$}+\text{CO$_2$}
\xrightarrow{R_{\text{H$_2$O}}}\text{H$_2$O}+\text{CO}.
\label{eq:water_production}
\end{equation}
Alumina is deposited at an (unknown) 
rate $R_{\text{Al$_2$O$_3$}}$ according to 
\begin{equation}
2\text{AlCl$_3$}+3\text{H$_2$O}
\xrightarrow{R_{\text{Al$_2$O$_3$}}}
\text{\alumina}+6\text{HCl}.
\label{eq:CVDalumina}
\end{equation}

\begin{figure}
\begin{tabular}{cc}
\epsfig{file=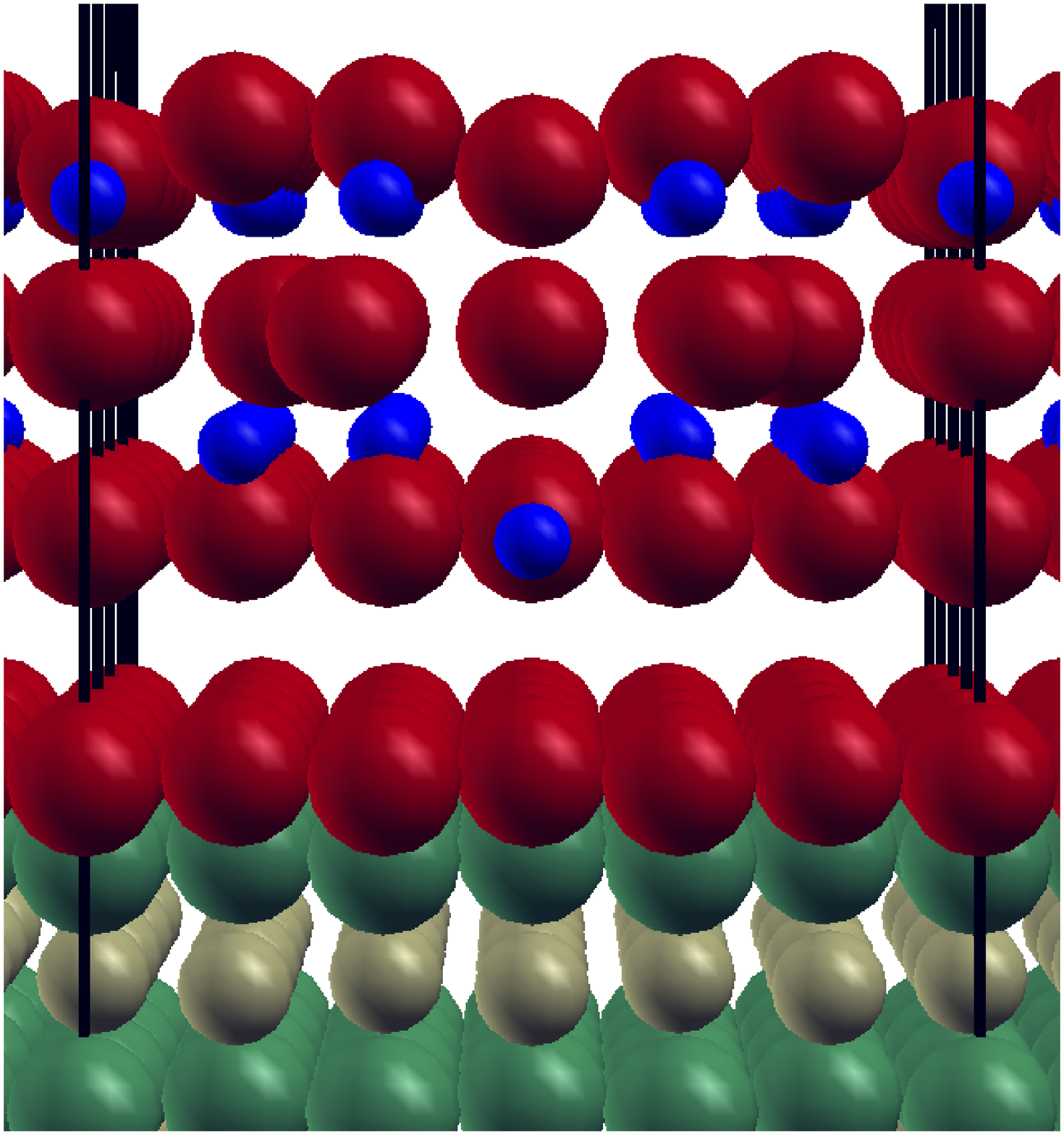,width=4cm}&
\epsfig{file=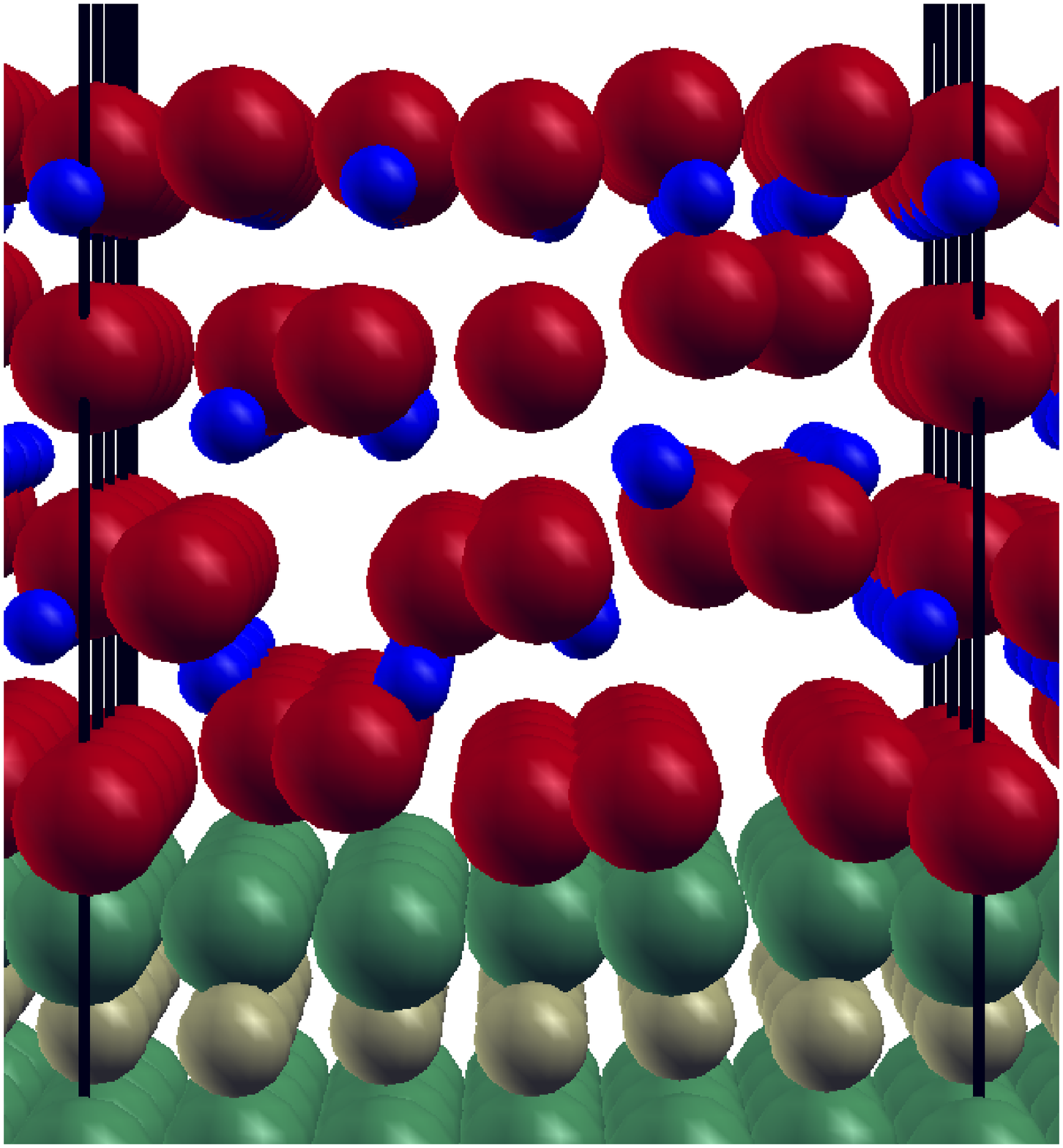,width=4cm}
\end{tabular}
\caption{
\label{fig:models}
Exfoliating (left panel) and wear-resistant alumina (right panel) films on TiC.
Color coding: Ti = dark green, C = light green, Al = blue,
O = red.
}
\end{figure}

\subsection{\textit{Ab initio} structure search for thin alumina films   
and TiC/alumina interface models} 
In Refs.~\cite{AluminaStructureSearch,Sead,IOPConf}  we have (in collaboration with others) 
presented  a  computational   strategy   for identifying energetically favorable geometries 
of thin-film alumina on a TiC substrate.
We have constructed alumina candidate configurations from a pool of structural motifs
as they exist in stable and metastable bulk alumina phases.
The motifs are characterized by stacking of the O layers and 
by the coordination of the Al ions.
We have considered alumina films of thicknesses up to four O layers. 
In addition, we allowed for off-stoichiometries in these films.

Each candidate configuration was structurally optimized by 
\textit{ab initio} total-energy and force calculations
and subsequent relaxation
(using standard quasi-Newton and conjugate gradient algorithms).
For a given thickness and stoichiometric composition,
the optimal geometry was identified as the one with lowest total energy
after structural relaxations.

Figure \ref{fig:models} details the atomic structure of two optimized alumina films
with a thickness of four O layers and different stoichiometric compositions.
These films are referred  to as exfoliating (left) and wear-resistant (right) films.
These terms refer to the adhesion properties of the two films
(exfoliating = no or very weak adhesion,
wear-resistant = strong adhesion).
Details concerning adhesion properties of these films are given elsewhere
\cite{AluminaAdhesion}.
For other thicknesses, the films possess similar structural (and adhesion) characteristics.
Focusing on the films in Fig.~\ref{fig:models} therefore does not pose a restriction of generality
on the presented discussion of the TiC/alumina interface.

The results of our (total-energy) structure search (for each class of Al$_M$O$_N$ films)
can be combined with the equilibrium thermodynamics  of Refs.~\cite{PhysRevB.62.4698,AIT_Scheffler,PhysRevB.70.024103}).
We find that this approach erroneously identifies the exfoliating film as thermodynamically stable \cite{AluminaStructureSearch}.
This incorrect equilibrium prediction applies to a wide range of temperatures and O$_2$ pressures
and is  in appearant conflict with wear-resistance
of TiC/alumina multilayer coatings \cite{Halvarsson1993177}.
Below we show that a more physical account of alumina growth emerges 
with our new nonequilibrium thermodynamics theory.

\subsection{Formation of excess atoms and free energies of reaction}
The films shown in Fig.~\ref{fig:models} 
can be divided into a stoichiometric part and an excess part.
For a general Al$_{M}$O$_{N}$ film, 
the number of stoichiometric units
in the stoichiometric part of the film is
$n_{\text{alumina}} = \min\left([M/2],[N/3]\right)$, where
$[x]$ is the largest integer smaller or equal to $x$.
The corresponding number of excess Al and O atoms is
$\Delta n_{\text{Al}}=N-2n_{\text{alumina}}$
and $\Delta n_{\text{O}}=M-3n_{\text{alumina}}$.

We assume that the deposition of the stoichiometric parts of the films is
described by  (\ref{eq:CVDalumina}).
Excess O can be deposited as 
\begin{subequations}
\begin{align}
\label{eq:ExcessAtoms}
\text{CO$_2$}\rightarrow \text{O$_{\text{exc}}$}+\text{CO}, \\
\text{H$_2$O}\rightarrow \text{O$_{\text{exc}}$}+\text{H$_2$},
\end{align}
\end{subequations}
depending on which reaction gives a lower free energy of reaction.
(Excess Al is not considered here.)
The free energy of reaction is defined accordingly,
\begin{align}
G_{\text{r}}^{\text{Al$_{M}$O$_{N}$}}
&=G_{\text{TiC/Al$_{M}$O$_{N}$}}
-G_{\text{TiC}}
+n_{\text{Al$_2$O$_3$}}\left(6  \mu_{\text{HCl}}
-2  \mu_{\text{AlCl$_3$}}
-3  \mu_{\text{H$_2$O}}\right)\nonumber\\
&+\Delta n_{\text{O}}\max\left[\mu_{\text{CO}}-\mu_{\text{CO$_2$}},
\mu_{\text{H$_2$}}-\mu_{\text{H$_2$O}}\right].
\label{eq:Hform}
\end{align}

\subsection{Approximations for evaluation of $G_r$}
We make standard  approximations for the free energy of solid 
and gaseous constituents.
The free energy of the TiC/alumina systems is approximated 
by their DFT total energies, that is, $G_{\text{solid}}\approx E_{\text{solid}}$.\footnote{
In fact, we correct the total energies of the films
by subtracting the strain energy of the stoichiometric
part of the film,
$E_{\text{film}}\rightarrow E_{\text{film}}-n_{\text{Al$_2$O$_3$}}\Delta_{\text{strain}}$,
where $\Delta_{\text{strain}}$ is the 
is difference  between the strained
(due to the expansion to the TiC lattice in the interface plane)
and the unstrained bulk alumina per stoichiometric unit.}
Vibrational effects are not considered \cite{AIT_Scheffler,AIT-DG_TiX}.
For gaseous constituents, we employ the ideal-gas approximation,
\begin{align}
\mu_i(T,p_i)=\epsilon_i+\Delta_i^0(T)+k_BT\ln(p_i/p^0).
\label{eq:mu}
\end{align}
Here $\epsilon_i$ is the DFT total energy \cite{molecules} of the gas phase species
(molecule),
and $k_B$ is  the Boltzmann constant.
$\Delta_i^0(T)$ is the temperature dependence
of $\mu_i$ at a fixed pressure $p^0$,
and available for $p^0=1$~atm
in thermochemical tables \cite{JANAF}.

\subsection{Rate-equation description of CVD of alumina}
The actual evaluation of the individual chemical potentials of the different
gaseous constituents in  the CVD chamber requires the specification of
the associated partial pressures.
In the actual fabrication process, these are, however, not directly controlled;
only the temperature and the total pressure are controlable.

We describe the CVD process in terms of rate equations for the individual
(ideal gas) pressures,
\begin{align}
\partial_tp_i
\propto
c_iR_{\text{S}}-\frac{p_i}{p}R_{\text{E}}
+\nu_i^{\text{H$_2$O}}R_{\text{H$_2$O}}
+\nu_i^{\text{Al$_2$O$_3$}}R_{\text{Al$_2$O$_3$}}.
\label{eq:RateEquation}
\end{align}
Here, $p_i=p_i(t)$ is the 
pressure of chemical species $i$ at time $t$ inside the reaction  chamber,
$p=p(t)=\sum_ip_i(t)$ is the corresponding total pressure,
$c_i$ is the concentration of the chemical species $i$ in the supply gas,
and $\nu_i^{\text{H$_2$O}}$ and $\nu_i^{\text{Al$_2$O$_3$}}$
are the stoichiometric coefficients\footnote{Here we use the standard conventions that 
stoichiometric coefficients are counted negative if a species is consumed
and positive if a species is produced in a reaction.}
of the the chemical species $i$ in  reaction
(\ref{eq:water_production}) and (\ref{eq:CVDalumina}), respectively
The rate at which the gas is supplied to (exhausted from) the chamber is
$R_{\text{S}}$ ($R_{\text{E}}$)
and the reaction rates for water production and alumina deposition
are $R_{\text{H$_2$O}}$ and $R_{\text{Al$_2$O$_3$}}$.

We use the resulting steady-state pressures 
\begin{align}
p_i=p
\frac{c_i+r_{\text{H$_2$O}}\nu_i^{\text{H$_2$O}}
+r_{\text{Al$_2$O$_3$}}\nu_i^{\text{Al$_2$O$_3$}}}
{1+r_{\text{Al$_2$O$_3$}}},
\label{eq:steady-state-pressures}
\end{align}
as input for the evaluation of the individual chemical potentials (\ref{eq:mu}).
In (\ref{eq:steady-state-pressures}) we have introduced
the  scaled reaction rates
$r_{\text{H$_2$O}}=R_{\text{H$_2$O}}/R_{\text{S}}$
and $r_{\text{Al$_2$O$_3$}}=R_{\text{Al$_2$O$_3$}}/R_{\text{S}}$.

\subsection{Limits on reaction rates}
The reaction rates $R_{\text{H$_2$O}}$ and $R_{\text{Al$_2$O$_3$}}$ cannot assume
any arbitrary value.
For example, we require that  $R_{\text{Al$_2$O$_3$}}\leq R_{\text{H$_2$O}}/3=R_{\text{Al$_2$O$_3$}}^{\text{max}}$,
since reaction (\ref{eq:CVDalumina}) requires three units of H$_2$O,
while reaction (\ref{eq:water_production}) produces only one.
Additional constraints on this
particular growth process and interface formation are 
discussed in Ref.~\cite{AluminaAdhesion}.

\section{Results}
\label{results}

\begin{figure}
\epsfig{file=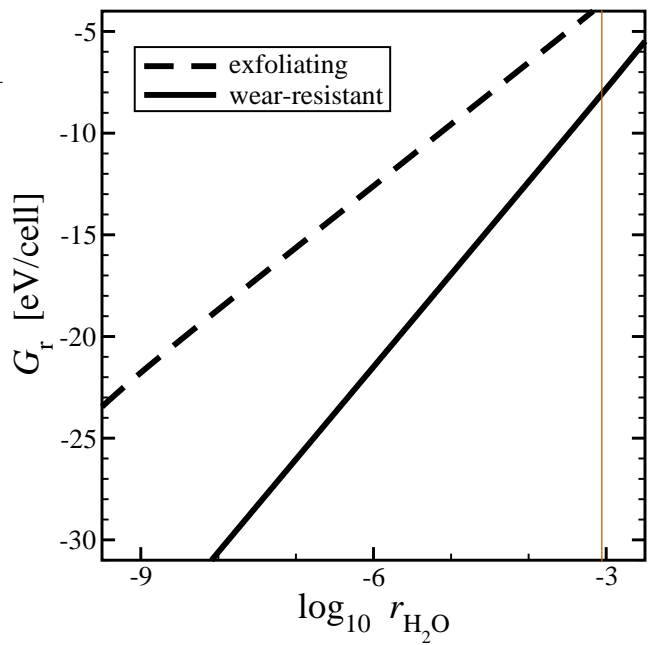,width=8.5cm}
\caption{
\label{fig:Gr}
Gibbs free energies of reaction  $G_{\text{r}}$ for exfoliating 
and wear-resistant alumina film (see Fig.~\ref{fig:models})
as functions of the scaled reaction rate $r_{\text{H$_2$O}}$.
The vertical line limits $r_{\text{H$_2$O}}$ to the right 
and corresponds to dynamic equilibrium in (\ref{eq:water_production}).
For simplicity, we assume $R_{\text{Al$_2$O$_3$}}=R_{\text{H$_2$O}}/3$.
The difference in $G_{\text{r}}$ is a predictor for the relative 
presence of exfoliating and wear-resitant films, see (\ref{eq:Pratio}).
}
\end{figure}

In Fig.~\ref{fig:Gr} we plot the free energies of reaction
for the two films shown in Fig.~\ref{fig:models} 
as  functions of the scaled reaction rate $r_{\text{H$_2$O}}$.
Deposition parameters (supply gas composition, total pressure,
and deposition temperature) as specified in Ref.~\cite{Ruppi200150}
have been used.
In this illustration of the approach we assume for simplicity
$R_{\text{Al$_2$O$_3$}}=R_{\text{H$_2$O}}/3$.
Furthermore, the scaled reaction rate $r_{\text{H$_2$O}}$
is limited to the right by the limit of dynamic equilibrium in 
(\ref{eq:water_production}), indicated by the vertical line.

We find that, over the entire range of the reaction rate $r_{\text{H$_2$O}}$,
it is more favorable to grow wear-resistant overlayers
than exfoliating overlayers.\footnote{
We notice that there is a third stoichiometric composition
(not discussed in this work)
that may be stabilized if $r_{\text{H$_2$O}}$ is sufficiently
far away from the dynamic equilibrium limit.
The corresponding structure has also wear-resistant properties \cite{AluminaAdhesion}.}
Qualitatively,
this result is independent of the choice of $R_{\text{Al$_2$O$_3$}}$.
We emphasize that this prediction  is consistent with industrial use 
of TiC/alumina as wear-resistant coating;
predictions based on an adaption of 
equilibrium-thermodynamics analysis \cite{PhysRevB.62.4698,AIT_Scheffler,PhysRevB.70.024103}
to this system \cite{AluminaStructureSearch} are not.

\section{Summary \& Conclusions}
\label{conclusions}
We have presented a novel computational scheme
to predict chemical compositions at interfaces
as they emerge in a growth process.
The scheme uses the Gibbs free energy of reaction associated
with the formation of interfaces with a specific composition
as predictor for their prevalence.
We explicitly account  for the growth conditions 
by rate-equation modeling of the deposition environment.

An earlier study of TiC/alumina interfaces \cite{AluminaStructureSearch} documented  
that the predicted composition at this interface is in conflict with the wear-resistance of multilayered TiC/alumina
when using an equilibrium-thermodynamics scheme \cite{PhysRevB.62.4698,AIT_Scheffler, PhysRevB.70.024103}.
Our results demonstrate that a careful account of deposition conditions
in interface modeling is crucial for understanding the
adhesion at  CVD TiC/alumina.

We expect that a similar analysis will be necessary 
also for other buried interfaces that form during a deposition
process in an environment that strongly differs 
from ambient conditions.
We emphasize the predictive power of the here-presented method,
adding to the DFT-based toolbox for accelerating innovation 
\cite{innovation}. 
In principle, it allows for the determination of deposition conditions
required to experimentally create interfaces
with a predetermined compositions.
We argue that combining the structure search method  of Ref.~\cite{AluminaStructureSearch}
and the new \textit{ab initio} thermodynamics
with, for example, a genetic algorithm
\cite{Finnis_GenAlg,Rohrer_GenAlg}
presents a valuable tool for characterization of surface and interface growth.

\section*{Acknowledgments}
We thank Carlo Ruberto and Gerald D.\ Mahan for valuable discussions.
We gratefully acknowledge support from the Swedish National Graduate School in Materials Science (NFSM),
from the Swedish Foundation for Strategic Research (SSF)
through ATOMICS,
from the Swedish Governmental Agency for Innovation Systems (VINNOVA),
from the Swedish Research Council (VR)
and from the Swedish National Infrastructure for Computing (SNIC).\\

\bibliographystyle{model1a-num-names}

\end{document}